\documentclass[12pt]{iopart}

\usepackage{graphicx}
\begin{document}

\title[Noise Line Identification in LIGO S6 and Virgo VSR2]{Noise Line Identification in LIGO S6 and Virgo VSR2}

\author{Michael Coughlin for the LIGO Scientific Collaboration and the Virgo Collaboration}

\address{Physics and Astronomy,
Carleton College,
Northfield, MN, 55057, USA}
\ead{coughlim@carleton.edu}
\begin{abstract}
An important goal for LIGO (the Laser Interferometer Gravitational-Wave Observatory) and Virgo is to find periodic sources of gravitational waves. The LIGO and Virgo detectors are sensitive to a variety of noise of non-astrophysical origin, such as instrumental artifacts and environmental disturbances. These artifacts make it difficult to know when a signal is due to a gravitational wave or noise. A continuous wave search algorithm, Fscan, and the calculation of the coherence between the gravitational wave channels and auxiliary channels has been developed to identify the source of noise lines. The programs analyze data from the gravitational wave channels as well as environmental sensors, searching for significant lines that appear in coincidence (using various thresholds and frequency windows) in the gravitational wave channel as well the environmental monitors. By this method, the source of powerful signals at specific frequencies in the gravitational wave channel caused by noise can be determined. Examples from LIGO's sixth science run, S6, and Virgo' second scientific run, VSR2, are presented.
\end{abstract}

\maketitle

\section{Introduction}

The general theory of relativity predicts that all accelerating objects with non-symmetric mass distributions produce gravitational waves (GW). GW presumably should be directly detectable when very massive objects such as black holes or neutron stars undergo acceleration. LIGO (the Laser Interferometer Gravitational-Wave Observatory) \cite{LIGO} and Virgo \cite{VIRGO} are some of the detectors searching for GW. These experiments seek to directly detect GW and use them to study astrophysical sources. They seek GW associated with the inspiral of binary neutron stars and black holes and the merger of these, GW burst from supernovae and gamma ray sources, periodic GW from nonaxisymmetric rotating or vibrating neutrons stars, and processes of the early universe which would produce a stochastic background of GW. \cite{LIGO}. 

In July 2009, LIGO commenced with its Sixth Science Run (S6), while Virgo started its Second Science Run (VSR2). The LIGO interferometers are located at Hanford, Washington (LHO) and Livingston, Louisiana (LLO), and Virgo is located in Cascina, Italy. These interferometers have resonant, Fabry-Perot arm cavities with light from lasers traveling down each of the arms \cite{LIGO,VIRGO}. Data from the interferometers is continuously collected for analysis at a variety of computer clusters operated by the LIGO Scientific Collaboration (LSC) and Virgo Collaborators.

Rapidly rotating neutron stars are the most promising sources of continuous-wave (CW) gravitational signals in the LIGO and Virgo operating frequency band \cite{WhitePaper}. There are approximately 200 known pulsars in our galaxy within the sensitive band of the interferometers, (those with spin frequencies greater than 20 Hz). They presumably emit gravitational radiation in a variety of ways, including elastic deformations, magnetic distortions, unstable r-mode oscillations, and free precession, all of which operate differently in accreting and non-accreting stars. The LSC and Virgo search for these sources, dividing their investigation into four broad categories: non-accreting known pulsars for which timing data are available, non-accreting known stars without timing data, non-accreting unknown stars, and accreting stars in known or unknown binary systems. There are indirect upper limits on continuous period GW emission that LIGO and Virgo have started to exceed \cite{Pulsar}. Most exciting, of course, would be the detection of these waves.

When LIGO and Virgo interferometers are ``locked'' and running stably with low noise, data are recorded in ``science time.'' In order to make detections of GW during episodes when the interferometer is in science time, it is necessary to understand when the data from the detectors are sufficiently ``clean'' or ``safe'' for observation of a GW. The LIGO and Virgo detectors are sensitive to a variety of noise sources of non-astrophysical origin, such as instrumental glitches, environmental disturbances, and mechanical resonances. Events not caused by GW in the data often produce significant effects in interferometers, as there is significant power in these signals. These instrumental and environmental artifacts make it difficult to identify a GW unambiguously.

As such, the development of techniques to identify and document non-astrophysical noise disturbances safely and effectively and thereby reduce their effect on searches for GW are needed. In order to accomplish this, a continuous wave search algorithm, Fscan, and the calculation of the coherence between the GW channels and auxiliary channels were developed as techniques for determining the frequencies at which the data coming into the LIGO detectors can be considered safe.

\section{Line Search Methods}

As discussed above, noise can affect GW detection. There is a concerted effort to locate and characterize the sources of noise, continuous and otherwise, in S6; a comprehensive description of the effort can be found here \cite{S6DetectorChar}. Lines due to noise usually occur at well defined frequencies in the detector frame (though some of these lines are rather broad, and others can wander). These lines can interfere with CW signal detection, even though CW signals have to be demodulated to correct for the Earth's motion relative to the source and intrinsic evolution of the source frequency. Thus, these signals can be masked by computer clocks and other digitized signals from the equipment running the detector as well as  periodic radio transmissions. They can also be masked by periodic environmentally triggered signals such as pumps or cooling fans, as well as vibrational modes of the suspension systems, and line noise due to the power supplies. Thus, the main goal of the project described here is to determine at what frequencies the data entering the detectors may be considered ``safe.'' The LSC and Virgo analyze the data from the GW channel and the other environmental sensors, look for sharp spectral lines at which there is a coincidence, and try to determine the cause of significant lines in the GW channel.

\subsection{Fscan}

\begin{figure}
\centering
\includegraphics[width=.6\textwidth]{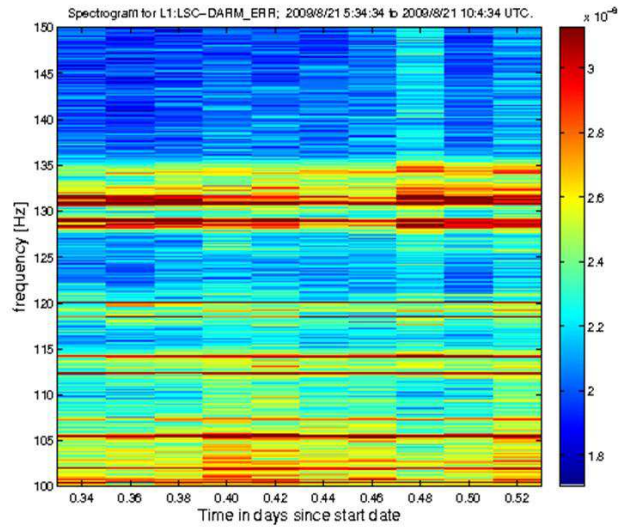}
\caption{A spectrogram based on the data from the FFT's. The intensity corresponds to the Fscan Power associated with each frequency over the given time interval.}
\end{figure}

The LSC has developed a program, written in the Matlab and C programming languages and run using Condor \cite{Condor}, which takes raw data from all of the sensors and Fast Fourier Transforms (FFT) the data, thereby representing them by their frequency components \cite{WhitePaper}. In this process, the power in each FFT in each frequency bin is normalized using a running median and then averaged over the FFTs, resulting in an ``Fscan Power'', with a typical frequency resolution of approximately 0.56 mHz. Using the FFT's, the program develops spectrograms as can be seen in Figure 1. The followup code, written in the Octave programming language, reads in the results of the FFT of the data and searches for significant lines that may appear. To do so, the program first removes all of the lines whose sources are known, including calibration lines, which are signals of single frequency continuously injected into the feedback control system to provide calibration; ``violin mode'', resonances in the wires supporting the mirrors; and power mains, including harmonics, which are the frequencies at which the AC power comes to the interferometers. Frequencies with power greater than a specified threshold are flagged as ``significant.'' The program assigns a ``line width'' to each frequency based on the measured strength of the FFT bins above or below it. It then looks for the first frequency bin above and below the line that has a power less than one half of the base frequency's power, and the line width is those two frequencies subtracted from one another. The program then compares the remaining above threshold frequencies from the GW channel and looks for coincident frequencies in the auxiliary channels using windows of 0.1, 0.01, and 0.001 Hz. If any two or more frequencies from the GW channel and the other channel fall within one of these windows, they are considered to be in coincidence and that information is recorded in a database.

The program subsequently plots Fscan Power vs. Frequency for the significant frequencies for the GW channel and the channel being compared, as can be seen in Figure 2. The goal of this is to give a visual representation of where frequencies may be clumping and also where the GW channel and the compared channel's frequencies appear coincident.

\begin{figure}
\centering
\includegraphics[width=.5\textwidth]{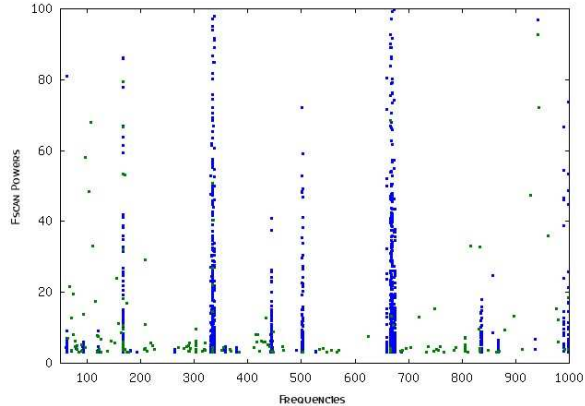}
\caption{A graph plotting the significant lines and their corresponding Fscan powers for the Virgo gravitational wave channel and an environmental monitoring channel (in this case, a seismometer on the injection bench).}
\end{figure}

The program next parses the lists of frequencies determined to be significant for that day and compares them to the frequencies in that same channel from the day before. The frequencies are compared with windows of the same size as in the comparison step explained above. A ``new line'' means that the given frequency appeared only in the data for the current day. An ``old line'' means that the given frequency presented in the data for both days. This approach allows those searching for the source of lines to track the presence of lines in specific channels. Both sporadic and steady lines are important to track.

In an effort to determine the origins of the significant lines in the GW channel, the program keeps track of data produced over the most recent seven calendar days. If a line in the GW channel comes up as significant every day over the past week, it is entered into a matrix, along with the number of times that the line appears in all of the other channels over the past seven days. The program applies a 0.001 Hz window to the matrix, in order to obtain the most accurate comparison. A more detailed matrix is assembled that also includes the frequencies appearing as significant in the GW channel as well as the channels appearing in coincidence. These matrices provide valuable diagnostics to scientists who are tracking specific lines.

Fscans for LHO, LLO, and Virgo calibrated strain h(t) data are created daily, weekly, and monthly. Doing so generates comparisons of significant lines in the GW channel between the sites. Lines appearing in two or three sites have the greatest potential for disrupting the GW search, and it is essential that these lines are understood so that they may be properly classified or discarded. In addition, the program is run on VSR2 data from a number of auxiliary channels, as well as the uncalibrated dark fringe output channel.

\subsection{Coherence}

The LSC has also developed a program to generate the basic coherence between the GW channel and all the available channels in the reduced data set (RDS), using a program written in the Python programming language. A follow-up program called Coherence then parses the information to facilitate investigation. Coherence first clusters lines above a predetermined threshold, the types of which are detailed below, to find significant lines and capture wide structures. The program next plots the coherence between the channels, as can be seen in Figure 4. Then, for each significant line in the GW channel, the program finds the channels for which there was significant coherence. Two sets of runs with different parameters are generated each week. One run uses a 1024 Hz detection band with a 1024 second window averaged over a week. The threshold for a significant line is defined as 15 times the theoretical $\sigma$, where the theoretical $\sigma$ is 1 over the number of time segments averaged for the coherence. The other run uses a 60 Hz detection band with a 60 second window with 100 averages. The threshold for significant lines is defined as coherence either greater than 35 times  theoretical $\sigma$ or greater than 0.95. These short runs are long enough to eliminate most of the false alarms and yet short enough to capture wandering lines like mechanical resonances.

\begin{figure}
\centering
\includegraphics[width=.6\textwidth]{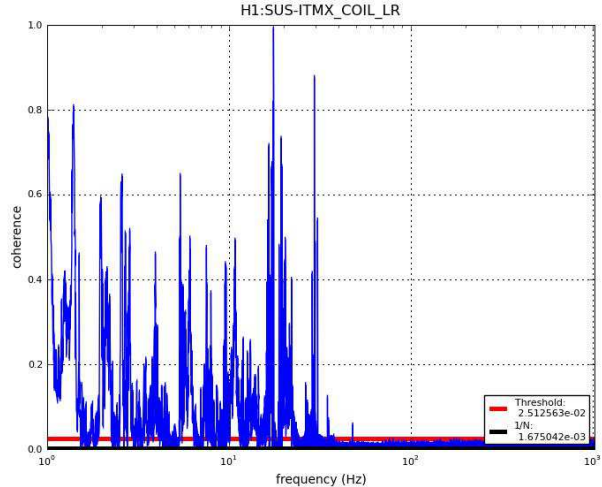}
\caption{A plot of the coherence vs. frequency between the GW channel and an example channel.}
\end{figure}

To help understand the results, the output of both Fscan and Coherence is stored in a database, and a web interface, developed to query results, is employed. Using this web interface, researchers can easily find, for example, coincident coherence lines in question.

\subsection{Line Searching}

In general, when searching for a line, the LSC and Virgo first look at the comparisons between the H1/L1/Virgo data for significant lines. If they see an unidentified significant line in coincidence among any combination of two or all three detectors, they will then search the Fscans for the environmental channels at LHO, LLO, and Virgo in an attempt to find the source of the lines. There are several other methods the LSC and Virgo use in its line searches \cite{Lines}. First of all, the LSC and Virgo report lines from search pipelines, such as PowerFlux \cite{PowerFlux} and Einstein@Home \cite{Einstein}, that are contaminating those searches and need to be identified. For troublesome frequencies like these, they create wiki pages, modifiable by all members, that contain information such as environmental channels which Fscan and other methods see in coincidence with that line. Direct comparisons between Fscan and Coherence also provide information about the lines. For example, the observation of a noise line in the Fscan and coherence results for auxiliary channels can help to identify the physical location of the noise, and whether its cause is likely to be due to electromagnetic, acoustic, mechanical, or seismic origin.

\section{Results}

\begin{figure}
\centering
\includegraphics[width=.6\textwidth]{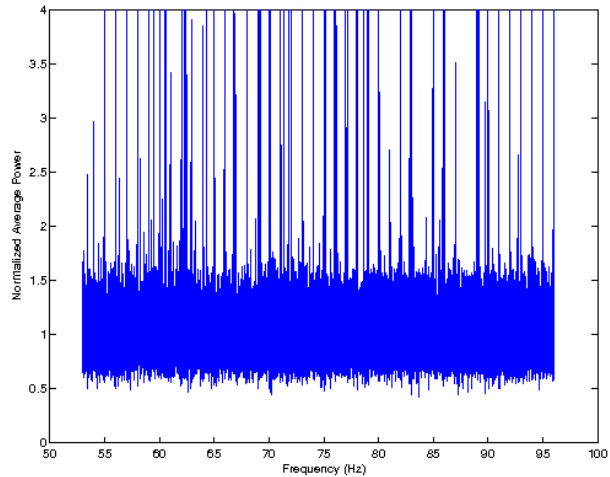}
\caption{A normalized spectrum plot for a Virgo environmental monitoring channel, in this case a magnetometer at the West end of Virgo.}
\end{figure}

Using the tools discussed above, the LSC and Virgo has flagged many frequency bands in the GW channels at all three sites that are, at present, of unknown origin. At both Hanford and Livingston, they have found that there are a large number of significant frequencies showing up daily which are integers, especially 2 Hz harmonics. This is unusual because only 1 out of 2000 frequencies analyzed are integers. Multiples of 16 Hz are suspected to arise within the data acquisition system, but there are many non-multiples of 16 Hz as well, and these will continue to be studied through comparisons between the two sites. Currently, this is attributed to recently installed digital electronics, but the exact source has yet to be found. 

In a comparison between LHO and LLO, aside from the integer frequencies, 54.496111 Hz and its harmonic 108.992222 Hz showed up as significant at both sites. Scientists at the observatories tracked the cause of the frequencies to the VME CPU's at the observatories. Such investigations will continue in seeking to identify lines found by Fscan, Coherence, and other tools. Similarly, the source of an 158 Hz line was found to be caused by a Foundry Ethernet switch using coincidence between the GW channel and three rack magnetometer channels in one of the buildings at Hanford.

At Virgo, as can be seen in Figure 4, a number of 10Hz harmonics show up strong in the Fscan output for not only the dark fringe channel, but also for a number of the environmental monitoring channels (especially the magnetometers). Similar to the LIGO observatories, a number of other integer frequencies show up significantly as well.

\section{Conclusion}

The most pressing goal for the near future is to identify the source of the many remaining unidentified noise lines. A related goal is to begin flagging ``wandering lines,'' that move between frequencies over the course of a continuous 24 hour period. With the current method, these frequencies get washed out due to the averaging technique.
	
Further work also includes examining weekly and monthly Fscans to answer a variety of questions such as: Where is the source of some of the lines that appear only over longer periods of averaging?	What type of lines only appear when different lengths of averaging are employed? A final project includes finding a reasonable way to compare the coherence and Fscan results. There are many lines that Coherence ``sees,'' but Fscan does not, and vice versa. This appears to happen because the coherence between the GW channel and an auxiliary channel can bring a noise line above the background while sometimes the noise line itself is not visible in the Fscan results. Conversely, noise lines seen in Fscan can sometimes move slightly, destroying the coherence with the GW channel and washing that line away in the coherence results. Understanding the source of the different lines are expected to advance efforts to understand more clearly why one method might ``see'' a line while another does not.

\ack

This project is funded by NSF Grant PHY-0854790 and the Kolenkow-Reitz Fund at Carleton. The authors gratefully acknowledge the support of the United States National Science Foundation for the construction and operation of the LIGO Laboratory, the Science and Technology Facilities Council of the United Kingdom, the Max-Planck-Society, and the State of Niedersachsen/Germany for support of the construction and operation of the GEO600 detector, and the Italian Istituto Nazionale di Fisica Nucleare and the French Centre National de la Recherche Scientifique for the construction and operation of the Virgo detector. The authors also gratefully acknowledge the support of the research by these agencies and by the Australian Research Council, the Council of Scientific and Industrial Research of India, the Istituto Nazionale di Fisica Nucleare of Italy, the Spanish Ministerio de Educaci\'on y Ciencia, the Conselleria d'Economia Hisenda i Innovaci\'o of the Govern de les Illes Balears, the Foundation for Fundamental Research on Matter supported by the Netherlands Organisation for Scientific Research, the Polish Ministry of Science and Higher Education, the FOCUS Programme of Foundation for Polish Science, the Royal Society, the Scottish Funding Council, the Scottish Universities Physics Alliance, The National Aeronautics and Space Administration, the Carnegie Trust, the Leverhulme Trust, the David and Lucile Packard Foundation, the Research Corporation, and the Alfred P. Sloan Foundation.


\section*{References}
\renewcommand\refname{Bibliography}
\bibliographystyle{unsrt}
\bibliography{FscanRef}

\begin{thebibliography}{1}

\bibitem{LIGO}
B.~Abbott et~al.
\newblock {LIGO: The Laser Interferometer Gravitational-Wave Observatory}.
\newblock {\em Reports on Progress in Physics}, 72, 2009.

\bibitem{VIRGO}
F.~Acernese et~al.
\newblock {Status of Virgo}.
\newblock {\em Classical and Quantum Gravity}, 25, 2008.

\bibitem{WhitePaper}
LIGO~Scientific Collaboration and Virgo Collaboration.
\newblock {The LSC-Virgo white paper on gravitational wave data analysis.
  Science goals, status and plans, priorities}.
\newblock {\em http://www.ligo.caltech.edu/docs/T/T080278-00.pdf}, 2009.

\bibitem{Pulsar}
Benjamin Owen.
\newblock {Detectability of periodic gravitational waves by initial
  interferometers}.
\newblock {\em Classical and Quantum Gravity}, 23, 2006.

\bibitem{S6DetectorChar}
Christensen~N (for~the LIGO Scientific~Collaboration and the
  Virgo~Collaboration).
\newblock {LIGO S6 Detector Characterization Studies. Submitted to these
  GWDAW14 proceedings, LIGO Document P1000045}.
\newblock 2010.

\bibitem{Condor}
Douglas Thain, Todd Tannenbaum, and Miron Livny.
\newblock {Distributed Computing in Practice: The Condor Experience}.
\newblock {\em Concurrency and Computation: Practice and Experience}, 17(2-4),
  2005.

\bibitem{Lines}
F.~Acernese et~al.
\newblock {Analysis of noise lines in the Virgo C7 data}.
\newblock {\em Classical and Quantum Gravity}, 24, 2007.

\bibitem{PowerFlux}
B.~Abbott et~al.
\newblock {All-Sky LIGO Search for Periodic Gravitational Waves in the Early
  Fifth-Science-Run Data}.
\newblock {\em Phys. Rev. Lett.}, 102(11), 2009.

\bibitem{Einstein}
B.~Abbott et~al.
\newblock {Einstein@Home search for periodic gravitational waves in LIGO S4
  data}.
\newblock {\em Phys. Rev. D}, D79, 2009.

\end{thebibliography}
\end{document}